\documentstyle{article}
\begin{document}
\begin{center}
{ \bf  Lamb shift and Stark effect  in simultaneous space-space and momentum-momentum  noncommutative quantum mechanics and $\theta$-deformed $su(2)$ algebra.}
\end{center}
\begin{center}

\textbf{S. A. Alavi}\\

\textit{Department of Physics, Sabzevar University of Tarbiat Moallem  , P. O. Box 397,
  Sabzevar, Iran }\\
\textit{Sabzevar House of Physics, Javan-Sara, Asrar Avenue, Sabzevar, Iran.}\\

\textit{Email: alavi@sttu.ac.ir, alialavi@fastmail.us}
 \end{center}


\emph{ We study the spectrum of Hydrogen atom, Lamb shift and Stark effect in the framework of simultaneous space-space and momentum-momentum (s-s , p-p) noncommutative quantum mechanics. The results show that the widths of Lamb shift due to noncommutativity is bigger than the one presented in [1]. We also study the algebras of abservables of systems of identical particles in s-s , p-p noncommutative quantum mechanics. We intoduce $\theta$-deformed $su(2)$ algebra.}\\

 \textbf{Introduction.}\\ 
 
It is generally believed that the picture of space-time as a manifold should break down at very short 
distances of the order of the  Planck length. Field theories on noncommutative spaces may play an 
important role in unraveling the properties of nature at the Planck scale. 
The study on noncommutative  spaces is much important for 
understanding phenomena at short distances beyond the present test of QED. It has been shown that the noncommutative geometry naturally appears in string theory with a non zero antisymmetric B-field [8].
Besides the string theory arguments the noncommutative field theories by themselves are very interesting. 
In recent years there have been a lot of work devoted to the study of NCFT's(or NCQM) and possible 
experimental consequences of extensions of the standard formalism to noncommutative one (see e.g.[1-25]).\\ 
In field theories the noncommutativity is introduced by replacing the
standard product by the star product. 
NCQM is formulated in the same way as the standard quantum mechanics SQM
(quantum mechanics in commutative spaces), that is in terms of the
same dynamical variables represented by operators in a Hilbert
space and a state vector that evolves according to the
Schroedinger equation :
\begin{equation}
i\frac{d}{dt}|\psi>=H_{nc}|\psi> ,
\end{equation}
we have taken into account $\hbar=1$. $H_{nc}\equiv H_{\theta}$
denotes the Hamiltonian for a given system in the noncommutative
space. In the literatures two
approaches have been considered  for constructing the NCQM :\\
a) $H_{\theta}=H$, so that the only difference between SQM and
NCQM is the presence of a nonzero $\theta$ in the commutator of
the position operators. \\
b) By deriving the Hamiltonian from the Moyal analog of the
standard Schroedinger equation :
\begin{equation}
i\frac{\partial}{\partial t}\psi(x,t)=H(p=\frac{1}{i}\nabla,x)\ast
\psi (x,t)\equiv H_{\theta}\psi(x,t) ,
\end{equation}
where $H(p,x)$ is the same Hamiltonian as in the standard theory,
and as we observe the $\theta$ - dependence enters now through the
star product [5]. In [6], it is shown that these two
approaches lead to the same physical theory.\\
In order to specify the phase space and the Hilbert space on which operators act one can take the Hilbert space to be 
exactly the same as the Hilbert space of Corresponding commutative systems [1]. There are different types of noncommutative theories [9]. For the phase space we consider both space-space and momentum -momentum 
noncommutativity. The space-space noncommutativity is inferred from the string theory [7,8]. The motivation for considering momentum-momentum noncommutativity are as follows :\\
a). To incorporate an additional background magnetic field [9,10].\\
b). To maintain Bose-Einstien statistics for systems of identical Bosons is constructed by generalizing one-particle 
quantum mechanics [12].\\
The noncommutative space can be realized by the coordinate operators satisfying :
\begin{equation}
[\hat{x}_{i},\hat{x}_{j}]=i\zeta^{-2}\Lambda^{-2}_{NC}d\theta_{ij}\hspace{1.cm}[\hat{x}_{i},\hat{p}_{j}]=i\delta
_{ij},\hspace{1.cm}[\hat{p}_{i},\hat{p}_{j}]=0,
\end{equation}

where $\theta_{ij}=\epsilon_{ijk}\theta_{k}$, is the noncommutativity parameter. $\Lambda_{NC}$  is the NC energy scale, and $d$ is a constant frame-independent dimensionless parameter. The scaling factor $\zeta$ will be defined later. In this paper we put $\theta_{3}=\theta$ and the rest of the $\theta$-components to zero , which can be done by a rotation or a redifinition of coordinates. The NC coordinates $\hat{x}$ and  momentum $\hat{p}$ in equ.(3), can be expressed in terms of commutative coordinates $x$  and $p$  as follows :
\begin{equation}
\hat{x}_{i}=x_{i}-\frac{1}{2}\zeta^{-2}\Lambda^{-2}_{NC}d\theta_{ij}p_{j},\hspace{1.cm}\hat{p}_{i}=p_{i}
\end{equation}

where now $x$ and $p$  satisfy in usual canonical commutation relations :
\begin{equation}
[x_{i},x_{j}]=[p_{i},p_{j}]=0,\hspace{1.cm}[x_{i},p_{j}]=i\delta
_{ij},
\end{equation}

In the case of simultaneous space-space and momentum-momentum noncommutativity, the consistent NCQM algebra is :
\begin{equation}
[\hat{x}_{i},\hat{x}_{j}]=i\zeta^{-2}\Lambda^{-2}_{NC}d\theta_{ij}\hspace{1.cm}[\hat{p}_{i},\hat{p}_{j}]=i\zeta^{-2}\Lambda^{2}_{NC}d^{\prime}\theta_{ij}\hspace{1.cm}[\hat{x}_{i},\hat{p}_{j}]=i\delta_{ij},
\end{equation}
 $d^{\prime}$  is another constant frame-independent dimensionless parameter. The scaling factor $\zeta$ is defined as :
\begin{equation}
\zeta=(1+\frac{dd^{\prime}}{4})^{\frac{1}{2}}
\end{equation}

 The NC coordinates $\hat{x}$  and  momentum  $\hat{p}$ can be written in terms of usual coordinates $x$ and momentum $p$ :
\begin{equation}
\hat{x}_{i}=x_{i}-\frac{1}{2}\zeta^{-2}\Lambda^{-2}_{NC}d\theta_{ij}p_{j}\hspace{1.cm}
\hat{p}_{i}=p_{i}+\frac{1}{2}\zeta^{-2}\Lambda^{2}_{NC}d^{\prime}\theta_{ij}x_{j}
\end{equation}

where $x$ and $p$ satisfy in equ.(5). We know that all the two dimensional antisymmetric tensors can be represented 
by the unit two dimensional antisymmetric tensor $\epsilon_{ij}$. The difference of the tensorial forms of the $x-x$ and $p-p$ commutators are represented by different cofficients $d$ and $d^{\prime}$. The dimensional parameters $\Lambda^{-2}_{NC}$ and $\Lambda^{2}_{NC}$ guarantee the correct dimensional of the tensorial forms of $x-x$ and $p-p$ commutators.\\

2. \textbf{ algebras of abservables of systems of identical particles in s-s, p-p  noncommutative two dimensional spaces.  $\theta$-deformed algebras.}\\

Heisenberg quantization for systems of identical particles in commutative case has been studied in detail in [26]. In this section we apply the Heisenberg quantization to systems of two identical particles in s-s,p-p noncommutative two dimensional spaces. The three dimensional case can be analyzed along similar lines. We can
describe the two particles systems by relative coordinates in the
same way as in one dimensional case. We define complex quantities
$\hat{a}_{j\pm}$ as follows :
\begin{equation}
\hat{a}_{j\pm}=\sqrt{\frac{\mu\omega}{2}}(\hat{x}_{j}\pm \frac{i}{\mu\omega}\hat{p}_{j}),\hspace{2 cm}j=1,2.
\end{equation}
where $\hat{x}_{\ell}$ and $\hat{p}_{\ell}$ satisfy in equ.(6). \\
It is shown in [12] that to maintain Bose-Einstein statistics the basic assumption is that operators $\hat{a}^{\dagger}_{i}$ and $\hat{a}^{\dagger}_{j}$ are commuting. This requirement leads to a consistency condition  $d^{\prime}=\mu^{2}\omega^{2}\Lambda^{-4}_{NC}d$.Then we have:

\begin{equation}
[\hat{a}_{j+},\hat{a}_{k+}]=[\hat{a}_{j-},\hat{a}_{k-}]=0 
\end{equation}
\begin{equation}
[\hat{a}_{j-},\hat{a}_{k+}]=\delta_{jk}+i \beta\theta_{jk}  .
\end{equation}

 where $\beta=\zeta^{-2}\mu\omega\Lambda^{-2}_{NC}d$. The generalized one-dimensional observables $A_{j}$, $B_{j}$ and $C_{j}$ are :
\begin{equation}
\hat{A}_{j}=\frac{1}{4}(\hat{a}_{j+}\hat{a}_{j-}+\hat{a}_{j-}\hat{a}_{j+})\hspace{2 cm}
\hat{B}_{j\pm}=\hat{B}_{j}\pm i \hat{C}_{j}=\frac{1}{2}(\hat{a}_{j\pm})^{2}.
\end{equation}
In addition we have two-dimensional observables which are the real
and imaginary parts of :
\begin{equation}
\hat{D}_{\pm}=\hat{D}_{re}\pm i \hat{D}_{im}=a_{1\pm} a_{2\pm} \hspace{2 cm}
\hat{E}_{\pm}=\hat{E}_{re}\pm i \hat{E}_{im}=\hat{a}_{1\mp} \hat{a}_{2\pm} .
\end{equation}
There are two $sp(1,R)$ algebras $\hat{A}_{1}$, $\hat{B}_{1\pm}$ and $\hat{A}_{2}$,
$\hat{B}_{2\pm}$  :
\begin{equation}
[\hat{A}_{j},\hat{B}_{j\pm}]=\pm \hat{B}_{j\pm} \hspace{1.5
cm}[\hat{B}_{j-},\hat{B}_{j+}]=2\hat{A}_{j}\hspace{.5 cm} j=1,2 .
\end{equation}

There are also two other algebras, one $\theta$-deformed $sp(1,R)$ algebra :
\begin{equation}
[\hat{A}_{1}+\hat{A}_{2},D_{\pm}]=\pm \hat{D}_{\pm}, 
\end{equation}
\begin{equation}
[\hat{D}_{+},\hat{D}_{-}]=-2(\hat{A}_{1}+\hat{A}_{2})+\theta \beta L ,
\end{equation}
where $L$ is the relative angular momentum operator.\\
and one $\theta$-deformed $su(2)$ algebra :
\begin{equation}
[\hat{A}_{2}-\hat{A}_{1},\hat{E}_{\pm}]=\pm \hat{E}_{\pm}-\frac{i}{2}\theta\beta [2(\hat{A}_{1}+\hat{A}_{2})+1]
\end{equation}
\begin{equation}
[\hat{E}_{+},\hat{E}_{-}]=2(\hat{A}_{2}-\hat{A}_{1}) .
\end{equation}
where $N$ is the number operator. To make this deformed algebra more familiar we use Schwinger's model notation 
\begin{equation}
J_{+}=a_{2-}a_{1+}\hspace{2. cm}J_{-}=a_{1-}a_{2+}
\end{equation} 
\begin{equation}
J_{z}=\frac{1}{2}(a_{2-}a_{2+}-a_{1-}a_{1+})\hspace{1. cm}N=(a_{2-}a_{2+}+a_{1-}a_{1+})
\end{equation} 
one can easily show that :

\begin{equation}
A_{2}-A_{1}=\frac{1}{2}(a_{2-}a_{2+}-a_{1-}a_{1+})=J_{z}
\end{equation}
\begin{equation}
A_{2}+A_{1}=\frac{1}{2}(a_{2-}a_{2+}+a_{1-}a_{1+}-1)=\frac{1}{2}(N-1)
\end{equation}

Eqs.(17-22) lead us to the following $\theta$-deformed $su(2)$ algebra :

\begin{equation}
[J_{+},J_{-}]=2J_{z}
\end{equation}
\begin{equation}
[J_{z},J_{\pm}]=\pm J_{\pm}-i\frac{1}{2}\theta\beta N
\end{equation}
To study the representation of this algebra we introduce operators $j_{+}$ and $j_{-}$ as follows :
\begin{equation}
j_{+}=J_{+}-\frac{i}{2}\theta\beta N. 
\end{equation}
\begin{equation}
j_{-}=J_{-}-\frac{i}{2}\theta\beta N,\hspace{2. cm} j_{z}=J_{z}
\end{equation}
One can show that $j_{+}$, $j_{-}$ and $j_{z}$ satisfy the ordinary $su(2)$ algebra :
\begin{equation}
[j_{+},j_{-}]=2j_{z}
\end{equation}
\begin{equation}
[j_{z},j_{\pm}]=\pm j_{\pm}
\end{equation}
Now everything goes as usual because representations of $su(2)$ algebra are well known. 

The other commutation relations are as follows : 
\begin{equation}
[\hat{E}_{-},\hat{D}_{+}]=2\hat{B}_{1+}-i\theta \beta D_{+}. 
\end{equation}
\begin{equation}
[\hat{D}_{-},\hat{E}_{+}]=2\hat{B}_{1-}+i\theta \beta D_{-}.
\end{equation}
\begin{equation}
[\hat{E}_{+},\hat{D}_{+}]=2\hat{B}_{2+}+i\theta \beta D_{+}.
\end{equation}
\begin{equation}
[\hat{D}_{-},\hat{E}_{-}]=2\hat{B}_{2-}-i\theta \beta D_{-} ,
\end{equation}
\begin{equation}
[\hat{E}_{-},\hat{B}_{2+}]=\hat{D}_{+}.
\end{equation}
\begin{equation}
[\hat{E}_{+},\hat{B}_{1+}]=\hat{D}_{+}.
\end{equation}
\begin{equation}
[\hat{E}_{-},\hat{B}_{1-}]=-\hat{D}_{-}.
\end{equation}
\begin{equation}
[\hat{E}_{+},\hat{B}_{2-}]=-\hat{D}_{-}.
\end{equation}
\begin{equation}
[\hat{D}_{-},\hat{B}_{1+}]=\hat{E}_{-}-i\theta\beta  a_{1-}a_{1+}  ,
\end{equation}
\begin{equation}
[\hat{D}_{+},\hat{B}_{2-}]=-\hat{E}_{-}+i\theta \beta a_{2-}a_{2+}
\end{equation}
\begin{equation}
[\hat{D}_{-},\hat{B}_{2+}]=\hat{E} _{+}+i\theta \beta a_{2+}a_{2-} .
\end{equation}
\begin{equation}
[\hat{D}_{+},\hat{B}_{1-}]=-\hat{E} _{+}-i\theta \beta a_{1+}a_{1-} .
\end{equation}

\textbf{3. Hydrogen atom spectrum ,Lamb shift and Stark effect in simultaneous s-s,p-p noncommutative spaces.}\\
Now we study the spectrum of Hydrogen atom, Lamb shift and Stark effect in the case of simultaneous space-space and momentum-momentum noncommutativity and compare with the results presented in [1].\\
Using equ.(8) the kinetic and potential terms can be written as follows :  

\begin{equation}
\frac{\vec{\hat{P}}}{2m}=\frac{\vec{P}}{2m}-\frac{\beta}{2m}\vec{L}.\vec{\theta}+O(\theta^{2}).
\end{equation}

\begin{equation}
V(\hat{r})=V(r)-\frac{Z e^{2}\beta}{4r^{3}}\vec{L}.\vec{\theta}+O(\theta^{2}).
\end{equation}
using the fact that $L.\theta=L_{z}\theta$ and :

\begin{equation}
\langle \ \ell j j_{z}\mid L_{z}\mid \acute{\ell} j \acute{j_{z}}\rangle =j_{z}\left(1\mp \frac{1}{2\ell +1}\right)\delta_{\ell\acute{\ell}}\delta_{j_{z}\acute{j_{z}}},\hspace{1. cm}j=\ell\pm \frac{1}{2}
\end{equation}
the energy level shift by (40) and (41) become :
\begin{equation}
\Delta E^{H-atom}_{NC}=-\left[ \frac {m_{e}}{4}(Z\alpha)^{4}\frac{\theta}{\lambda^{2}_{e}}\beta f_{n,\ell}+\frac{\theta\beta}{2m}\right]j_{z}\left(1\mp \frac{1}{2\ell +1}\right)\delta_{\ell\acute{\ell}}\delta_{j_{z}\acute{j_{z}}}
\end{equation}

As it is observed and mentioned in [1], Lamb shift i.e. $2P_{\frac{1}{2}}\rightarrow 2S_{\frac{1}{2}}$ transition differs from the usual commutative case in which the shift depends only on the $\ell$ quantum number and all the corrections are due to the field theory loop effects. The Lamb shift for the simultaneous space-space and momentum-momentum noncommutative H-atom, besides the usual loop effects, depends on the $j_{z}$ quantum number. There is also a new channel which is opened because of noncommutativity : $2P^{-\frac{1}{2}}_{\frac{1}{2}}\rightarrow 2P^{\frac{1}{2}}_{\frac{1}{2}}$. The usual Lamb shift , $2P_{\frac{1}{2}}\rightarrow 2S_{\frac{1}{2}}$  is now split into two parts, $2P^{-\frac{1}{2}}_{\frac{1}{2}}\rightarrow 2S_{\frac{1}{2}}$ and $2P^{\frac{1}{2}}_{\frac{1}{2}}\rightarrow 2S_{\frac{1}{2}}$, which means that the simultaneous noncommutativity effects increase  the widths and split the Lamb shift line by a factor proportional to $\theta$, but the gap widths  between two parts is bigger than the case of single space-space noncommutativity(i.e. the results presented in [1]). only experiments can tell whether the space is really noncommutative or not, and in case it is, which is the non-commutative structure, simultaneous space-space and momentum-momentum or only space-space noncommutativity. \\ 
Now let us study the Stark effect in the case of simultaneous s-s and p-p noncommutativity. The potential energy of the atomic electron in an external electric field oriented along the $z$ axis is given by :
\begin{equation}
V_{Stark}=eEz+\frac{e\beta}{4}\left(\theta \times p\right)\cdot E +\frac{\beta}{2m}\left(\theta \times p\right)\cdot (r^{3}E)
\end{equation}
To the first order in perturbation theory the contribution to the Stark effect due to the second term is zero [1]. For the third term to the first order  we have :
\begin{equation}
\Delta^{NC}_{Stark}\propto (\vec{\theta}\times \vec{E})_{i}\left\langle  n\ell^{\prime}jj^{\prime}_{z}\left|p_{i}
r^{3}\right|n\ell jj_{z}\right\rangle\neq 0
\end{equation}
which  is different from the result presented in [1].\\

\textbf{Conclusion. }\\
In conclusion we have studied the spectrum of Hydrogen atom , Lamb shift and Stark effect in simultaneous space-space and momentum-momentum noncommutative spaces which are different from those in the case of space-space noncommutativity.  As we mentioned in the text only experiments can tell whether the space is really noncommutative or not, and in case it is, which is the noncommutative structure. We have also introduced new deformed $su(2)$ algebra which appears when one study Schwinger's model or quantization of systems of identical particles in simultaneous space-space and momentum-momentum noncommutative spaces.\\

\textbf{ Acknowledgments.}\\
I would like to thank  Prof. Jian-zu Zhang and M. M. Sheikh-Jabbari for their  careful reading of the manuscript and for their valuable comments. \\
  
\textbf{References. }\\
1. M. Chaichian, M. M. Sheikh-Jabbari and A. Tureanu, Phys. Rev.
   Lett. 86, 2716 (2001) and Eur.Phys.J. C36 (2004) 251.\\
2. N. Chair, M. A. Dalabeeh, hep-th/0409221.\\
3. A. E. F. Djemai, H. Smail, Commun. Theor. Phys. 41(2004) 837.\\
4. J. Douari, hep-th/0408150.\\ 
5. L. Mezincescu, hep-th/0007046.\\
6. O. Espinosa, P. Gaete, hep-th/0206066.\\
7. F. Ardalan, H. Arfaei and M. M. Sheikh-Jabbari, JHEP 9902, (1999) 016.\\
8. N. Seiberg, E. Witten, JHEP 9909 (1999) 032.\\
9. M. R. Douglas, N. A. Nekrasov, Rev. Mod. Phys 73 (2001) 977.\\
10. V. P. Nair, A. P. Polychronakos, Phy. Lett. B505 (2001) 267.\\
11. Y. Myung , Phys.Lett. B601 (2004) 1.\\
12. Jian-zu Zhang, Phys. Lett. B584 (2004) 204. \\
13. B. Durhuus, T. Jonsson, JHEP 0410 (2004) 050.\\\
14. Valentin V. Khoze, J. Levell , JHEP 0409 (2004) 019.\\
15. K. Fujikawa, Phys.Rev. D70 (2004) 085006.\\
16. A. Kokado, T. Okamura, T. Saito, Phys.Rev. D69 (2004) 125007.\\
17. A. H. Fatollahi, H. Mohammadzadeh, Eur.Phys.J. C36 (2004) 113-116.\\
18. B. Ydri, Mod.Phys.Lett. 19 (2004) 2205-2213.\\
19. C. Zachos, Mod.Phys.Lett. A19 (2004) 1483-1487.\\
20. T. A. Ivanova, O. Lechtenfeld, H. Mueller-Ebhardt, Mod.Phys.Lett. A19 (2004) 2419.\\
21. S. A. Alavi, Mod. Phys. Lett. A in press, hep-th/0412292 .\\
22. R. J. Szabo, Int.J.Mod.Phys. A19 (2004) 1837-1862.\\
23. P. Valtancoli, Int.J.Mod.Phys. A19 (2004) 4789-4812 and Int.J.Mod.Phys. A19 (2004) 4641.\\
24. S. A. Alavi, F. Nasseri, Int.J.Mod.Phys.A in press, astro-ph/0406477.\\
25. F. Nasseri, S. A. Alavi, Int.J.Mod.Phys.D in press, hep-th/0410259.\\
26. J. M. Leinaas, J. Myrheim, Int. J. Mod. Phys A8 (1993) 3649.\\   

\end{document}